\documentclass{aastex61}

\shortauthors{Xu et al.}

\begin{document}

\title{Enhancing Fault Diagnosis in GWAC: A Monitoring System for Telescope Arrays}

\correspondingauthor{Y. Xu}
\email{yxu@nao.cas.cn}

\author[0000-0002-8694-4672]{Y. Xu}
\affil{CAS Key Laboratory of Space Astronomy and Technology, National Astronomical Observatories, Chinese Academy of Sciences, Beijing, 100012, China}

\author{G. W. Li}
\affiliation{CAS Key Laboratory of Space Astronomy and Technology, National Astronomical Observatories, Chinese Academy of Sciences, Beijing, 100012, China}

\author{J. Wang}
\affiliation{CAS Key Laboratory of Space Astronomy and Technology, National Astronomical Observatories, Chinese Academy of Sciences, Beijing, 100012, China}

\author{L. P. Xin}
\affiliation{CAS Key Laboratory of Space Astronomy and Technology, National Astronomical Observatories, Chinese Academy of Sciences, Beijing, 100012, China}

\author{H. B. Cai}
\affiliation{CAS Key Laboratory of Space Astronomy and Technology, National Astronomical Observatories, Chinese Academy of Sciences, Beijing, 100012, China}

\author{X. H. Han}
\affiliation{CAS Key Laboratory of Space Astronomy and Technology, National Astronomical Observatories, Chinese Academy of Sciences, Beijing, 100012, China}

\author{X. M. Lu}
\affiliation{CAS Key Laboratory of Space Astronomy and Technology, National Astronomical Observatories, Chinese Academy of Sciences, Beijing, 100012, China}

\author{L. Huang}
\affiliation{CAS Key Laboratory of Space Astronomy and Technology, National Astronomical Observatories, Chinese Academy of Sciences, Beijing, 100012, China}

\author{J. Y. Wei}
\affiliation{CAS Key Laboratory of Space Astronomy and Technology, National Astronomical Observatories, Chinese Academy of Sciences, Beijing, 100012, China}
\affiliation{University of Chinese Academy of Sciences, Beijing, 100049, China}



\begin{abstract}

The Ground-based Wide-Angle Cameras array (GWAC) necessitates the integration of over 100 hardware devices, more than 100 servers, and upwards of 2500 software modules, all synchronized within a 3-second imaging cycle. However, the complexity of real-time and high concurrency processing of big data have historically resulted in a substantial failure rate, with estimated observation efficiency of less than 50\% in 2023. 
To address these challenges, we developed a  monitoring system aimed at enhancing fault diagnosis efficiency. 
The system features two innovative monitoring views: state evolution monitoring and transient lifecycle monitoring. These, combined with instantaneous state monitoring and key parameter monitoring views, create a comprehensive and holistic monitoring strategy.
This paper details the system's architecture, data collection methods, and the design philosophy of monitoring views. 
After a year of practical fault diagnostics, the system has demonstrated the ability to identify and localize faults within minutes, achieving fault localization speeds nearly ten times faster than traditional methods. Additionally, the system's design exhibits high generalizability, making them applicable to other telescope array systems.

\end{abstract}

\keywords{Automated telescopes,  Astronomy image processing, fault diagnosis, monitoring system}



\section{Introduction} \label{sec:intro}

The Ground-based Wide-Angle Cameras array (GWAC), one of the main ground-based facilities of the Chinese-French SVOM mission \citep{2016arXiv161006892W},  is to construct a fully automated, self-triggering, and self-follow-up transient survey system, and has completed the construction of the real-time autonomous detection for transients. GWAC has reported a series of scientific results, such as the independent detection of one high-energy burst, GRB 201223A \citep{2023NatAs...7..724X}, with a time scale of only 29 seconds; the discovery of more than 200 white-light flares \citep{2024ApJ...971..114L, 2023RAA....23a5016L}; the discovery of two cold star superflares with amplitudes reaching around 10th magnitude \citep{2024MNRAS.527.2232X, 2021ApJ...909..106X}; and long-term activity research on cold star flares \citep{2023ApJ...954..142L}. The GWAC consists of 10 mounts, 50 image sensors, 50 camera focusing decices, and over 100 supporting servers. The data processing pipeline covers multiple subsystems such as observation scheduling \citep{2021PASP..133f5001H}, observation control, automatic focusing [Huang, 2016], automatic guiding, real-time scientific data processing, automatic follow-up \citep{2020PASP..132e4502X} , and scientific result management \citep{2021RMxAC..53..174X}.

The daily operations of the GWAC rely on the coordination of various software and hardware modules, involving complex processes that result in a relatively high fault rate. The two primary challenges are as follows:
\begin{enumerate}
\item Hardware-Software Dependency: The hardware operations depend heavily on software feedback. For example, the mount's pointing correction relies on astrometric results from observed images, and maintaining image quality depends on image assessment results. These results are affected by varying weather conditions, introducing uncertainty in astrometric and image assessments. Since scientific data processing also depends on these feedback results, the overall feedback loop is long and lacks reliability.

\item Complex Software Architecture: The software architecture is intricate, requiring high concurrency and real-time processing. The data processing pipeline must address various tasks such as observation planning, image quality assessment, template catalog generation, hardware feedback, cross-matching with multiple catalogs, and automated follow-up of transient. Each image processed requires over 50 software modules, and with all cameras operating simultaneously, more than 2,500 module invocations occur per image cycle. The interdependencies between these modules mean that a single failure can disrupt the entire process, reducing system robustness.
\end{enumerate}

When the number of telescopes is small, daily malfunctions are less frequent and have a minimal impact on observation efficiency. However, as the telescope number increases, malfunctions become inevitable, often surpassing the diagnostic and repair capacity of the maintenance staff. This results in prolonged operation with faults and a significant drop in system efficiency. Therefore, GWAC urgently needs an efficient monitoring system to continuously track the system's status, provide fault diagnosis alerts, and guide maintenance staff in quickly resolving issues, ultimately enhancing operational efficiency.

Current monitoring solutions are inadequate for GWAC's needs due to its complexity, high concurrency, and real-time requirements. For instance, the Cherenkov Telescope Array monitoring system \citep{2022icrc.confE.700C} primarily tracks hardware degradation to prevent major failures. LAMOST \citep{2021MNRAS.500..388H} focuses on real-time assessments of guiding system performance, focal surface defocus, submirror performance, and active optics system performance. Similarly, radio astronomy projects like SKA \citep{2016SPIE.9913E..3SD} deploy monitoring software mainly for hardware health status. These systems emphasize hardware health or observation efficiency but are insufficient for GWAC. GWAC, being a complex system integrating high-concurrency data processing and precision mechanical control, requires a monitoring system that offers real-time hardware status tracking and comprehensive software pipeline monitoring.

The GWAC monitoring system oversees the full running status of all telescopes and presents monitoring data through diverse views. To ensure comprehensive fault coverage, the system employs various data collection methods, including hardware status, real-time pipeline status, and key image parameters. During the collaboration of hardware and software, the system generates massive amounts of raw monitoring data, surpassing manual analysis capabilities. By integrating and abstracting the raw data, the system provides multidimensional monitoring views that simplify data interpretation and reduce reliance on the experience and skills of maintenance personnel. Additionally, it enables deeper insights into internal operations and evolutionary processes, supporting manual fault diagnosis and laying the groundwork for future automated fault detection.

This paper introduces a monitoring system for GWAC, which designs a diversified monitoring data collection and visualization scheme. The structure of the paper is as follows: Section \ref{sec:system} describes the system architecture and its relationship with existing GWAC pipelines; Section 3 details the design and implementation of the system, including the construction of monitoring views, database design, and system implementation; Section 4 presents fault diagnosis cases based on the monitoring views; and the final section summarizes and provides an outlook for the proposed monitoring system.

\section{System Architecture} \label{sec:system}

\begin{figure}
\plotone{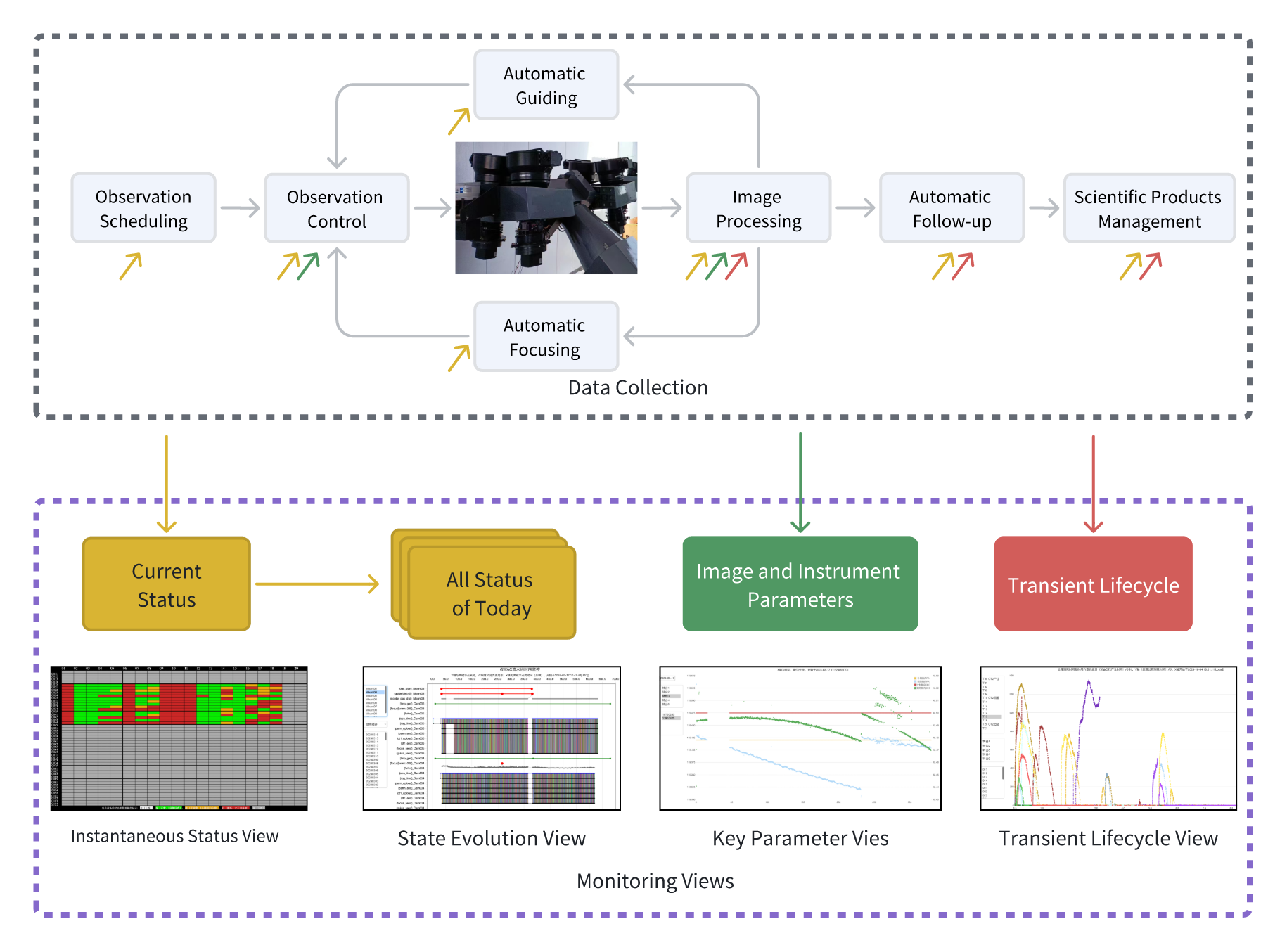}
\caption{The architecture of GWAC monitoring system. The system consists of two main components: data collection and monitoring views. Data collection is integrated into the existing GWAC pipeline to collect data and generate four monitoring views.\label{fig:sysArch}}
\end{figure}

To enhance the efficiency of fault detection and diagnosis in the GWAC system, we developed a monitoring system integrated into the existing GWAC pipeline. As illustrated in Figure \ref{fig:sysArch}, the monitoring system comprises two main components: data collection and monitoring views.

\subsection{Data Collection} \label{subsec:dataCollection}

The data collection component, indicated by the gray dashed box in Figure \ref{fig:sysArch}, aggregates data from seven subsystems in the GWAC data processing pipeline: the Observation Scheduling System, Observation Control System, Automatic Guiding System, Automatic Focusing System, Image Processing System, Automatic Follow-up System, and Scientific Results Management System. The collected data falls into three categories: key module invocation times, image and instrument parameters, and transient processing information, represented by the orange, green, and red arrows, respectively. The presence of these colored arrows next to a subsystem’s box in the figure indicates the type of data collected for that subsystem. For instance, the invocation time of key modules is recorded across all seven subsystems. The three data types are described as follows:

\begin{enumerate}
\item Key Module Invocation Time:
The seven GWAC subsystems comprise over 50 software modules. Monitoring all modules would produce excessive data, complicating management and visualization. Thus, only key modules are selected for monitoring. For each key module, the invocation start time is recorded to indicate its activity status. The specific key modules will be introduced in Section 3 on Instantaneous State Monitoring.
\item Image and Instrument Parameter:
In the GWAC data processing pipeline, each observation image is subjected to various analyses, including image quality, pointing accuracy, alignment precision, and target counting. The outcomes of these analyses collectively form the image parameters. Additionally, instruments like the mount and camera generate state parameters (e.g., temperature, vacuum level, voltage, current), constituting the instrument parameters.
\item Transient Processing Information:
Transients are crucial scientific outputs of GWAC. Their processing timeliness significantly affects scientific results. This part records the start times of transient related key modules, from the discovery frame of transient to the trigger of follow-up observations, to track transient processing.
\end{enumerate}

\subsection{Monitoring views} \label{subsec:monitorViews}

The monitoring views component (purple dashed box in Figure 1) abstracts the collected data into four views: instantaneous status, state evolution, key parameter, and transient lifecycle monitoring view. These views will be elaborated in Section 3. A brief overview follows:

\begin{enumerate}
\item Instantaneous State Monitoring:
This view tracks real-time system status, providing immediate feedback on system health. It alerts staff in case of malfunctions for prompt diagnosis and resolution. The view includes real-time camera previews and key module status displays.
\item State Evolution Monitoring:
This view shows the evolution of key module states over time, similar to the integration of instantaneous state monitoring with a timeline.
\item Key Parameter Monitoring:
This view primarily monitors the parameters related to images, mounts, and cameras, showing the trend of these parameters over time.
\item Transient Lifecycle Monitoring:
This view visualizes the lifecycle of transients, from the discovery frame to the trigger for follow-up observations. It includes key timestamps: detection frame of transient candidate, start and end of identification process, start and end of follow-up observation, providing insights into the processing pipeline's normalcy.
\end{enumerate}

\section{System Design and Implementation} \label{sec:systemDesign}

\subsection{Monitoring View Design} \label{subsec:monitorViewDesign}
Managing over 100 custom hardware devices, 100+ servers, and 2,500+ software modules on limited screen space poses significant challenges in presenting status and parameter data efficiently. This requires a well-structured user interface, a thorough understanding of monitoring parameters, and abstract representations. To enhance fault analysis efficiency, the number of monitoring pages should be limited, with each page designed for clarity. Excessive page switching can result in information loss and longer fault analysis times, necessitating a compact information presentation scheme. Based on these principles, we developed four monitoring views to provide a comprehensive operational overview of the GWAC system, facilitating quick fault detection and localization.

\subsubsection{Instantaneous Status Monitoring}

This view offers early warnings and an at-a-glance summary of the system's current health, including camera observation images monitoring and key module status monitoring view. The specific UI design details are shown in Figure 2. The following sections introduce these two parts separately.

\begin{enumerate}
\item Camera Observation Image Monitoring 

This page displays real-time images from each camera to check focus, camera status, weather conditions, etc. The design challenge is to maintain clarity when displaying images from 50 cameras simultaneously. We use a combination of thumbnails and a high-resolution image carousel (Figure 2, left). Thumbnails are displayed on the sides for quick browsing, while high-resolution images are shown in the center carousel, enabling users to click thumbnails for detailed views.

\item Key Module Instantaneous Status Monitoring 

This page tracks the real-time status of each camera's data processing pipeline, involving over 50 modules. Due to space constraints, only key modules are monitored and displayed graphically (Figure 2, right). Each row corresponds to a camera, and columns represent key modules. States are color-coded: white (online), green (normal), orange (warning), red (fault), and gray (offline). This layout efficiently conveys critical information, aiding quick fault diagnosis.

\end{enumerate}

\begin{figure}
\gridline{
  \includegraphics[height=4cm]{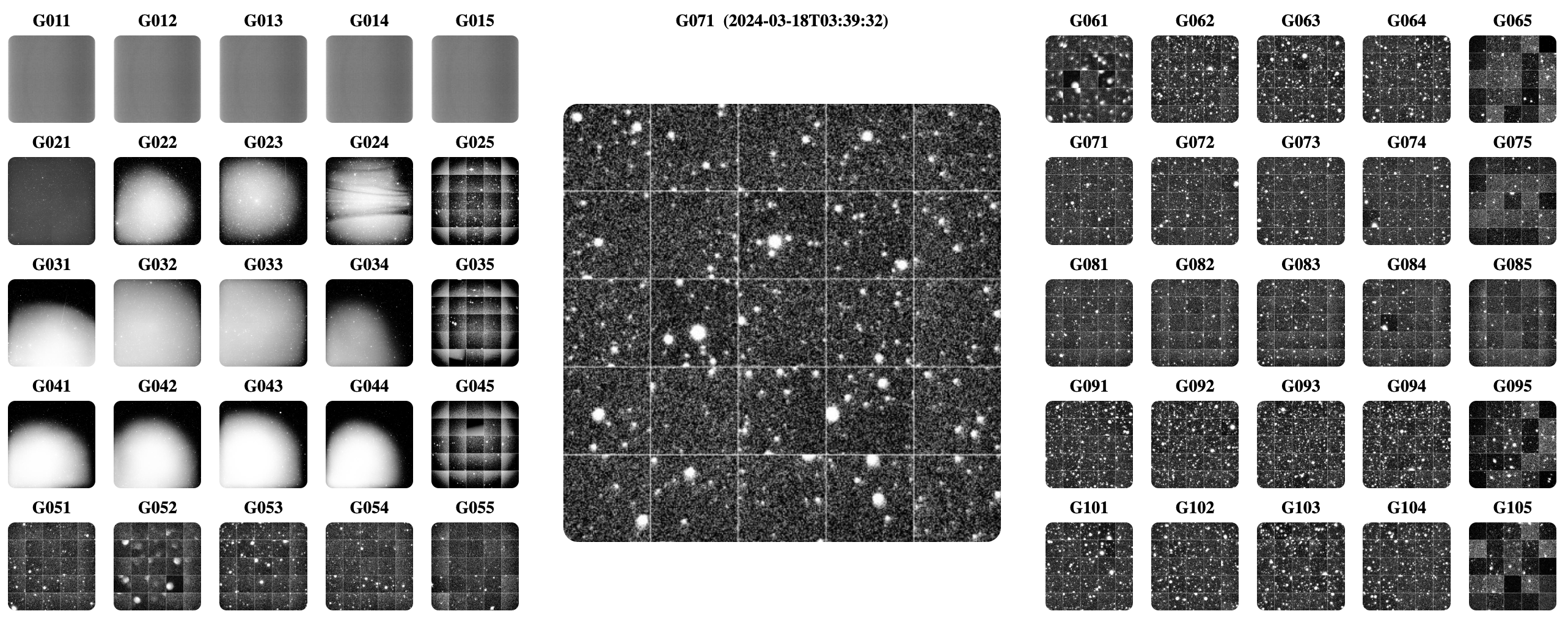}
  \includegraphics[height=4cm]{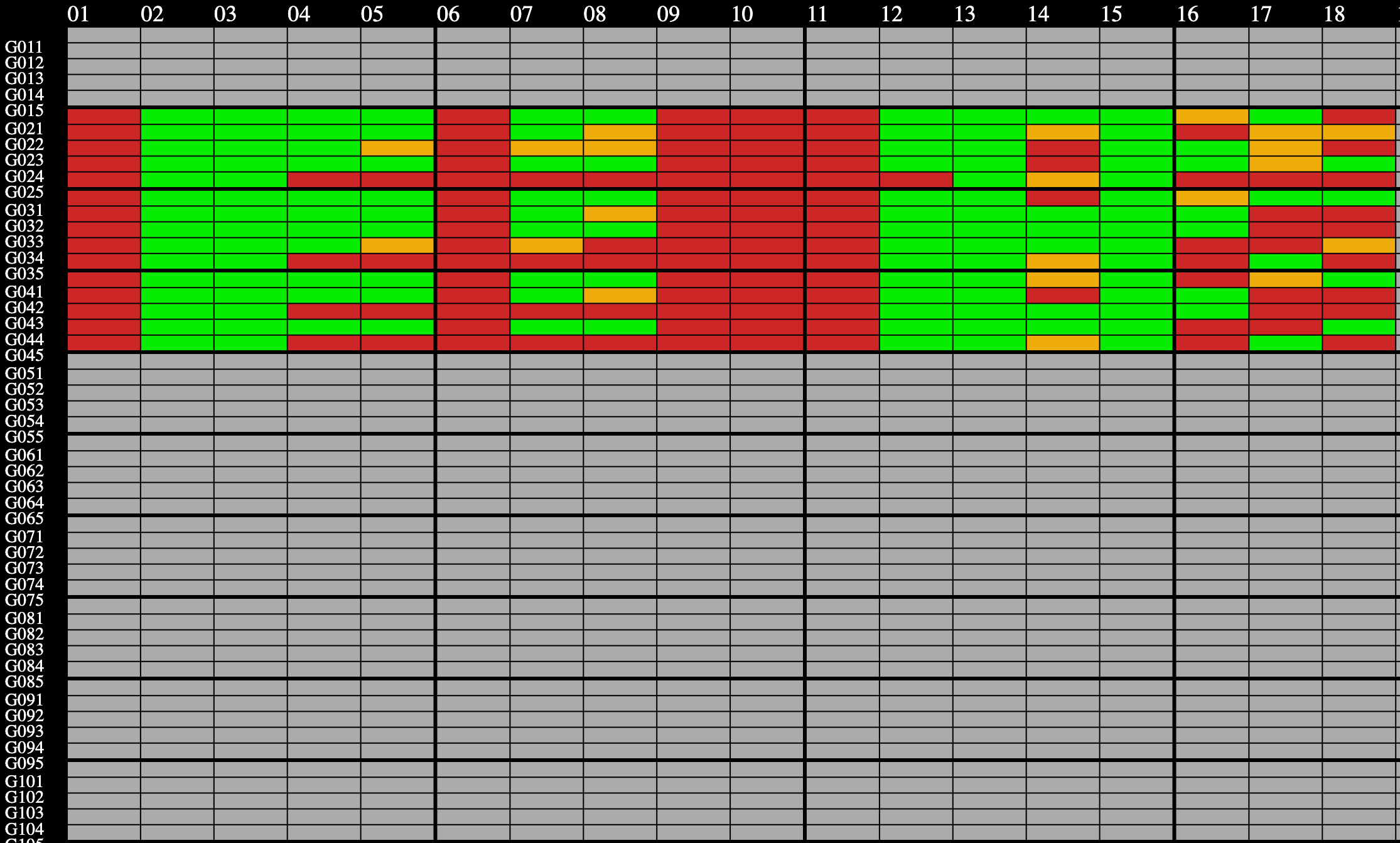}
}
\caption{Instantaneous Status Monitoring view. The left image shows a real-time preview of the camera observation, while the right image displays the instantaneous status monitoring of key modules.\label{fig:instantStatus}}
\end{figure}

\subsubsection{State Evolution Monitoring}

State evolution monitoring addresses the continuous changes in the system during observation, highlighting complex fault phenomena crucial for diagnosis. Traditional methods require specialized tools, which are time-consuming and demand expertise. 
We designed a new view to visualize the dynamic system state, showing changes in mount pointing, image quality, and image lifecycle, as shown in Figure 3. 
The UI consists of a control area for mount selection, data loading selection, and observation dates selection, alongside a chart showing parameter changes over time.
\begin{itemize}
\item X-axis: Time in minutes, relative to the observation start.
\item Y-axis: Parameters such as mount properties (observation plans, guiding actions, pointing errors), camera properties (template image creation, focusing, image quality like FWHM) and key module invocation time in image processing pipeline. The five cameras' properties on one mount are arranged in order.
\item Event Association: Vertical lines link key modules of each image, indicating different types of faults.
\end{itemize}

\begin{figure}
\plotone{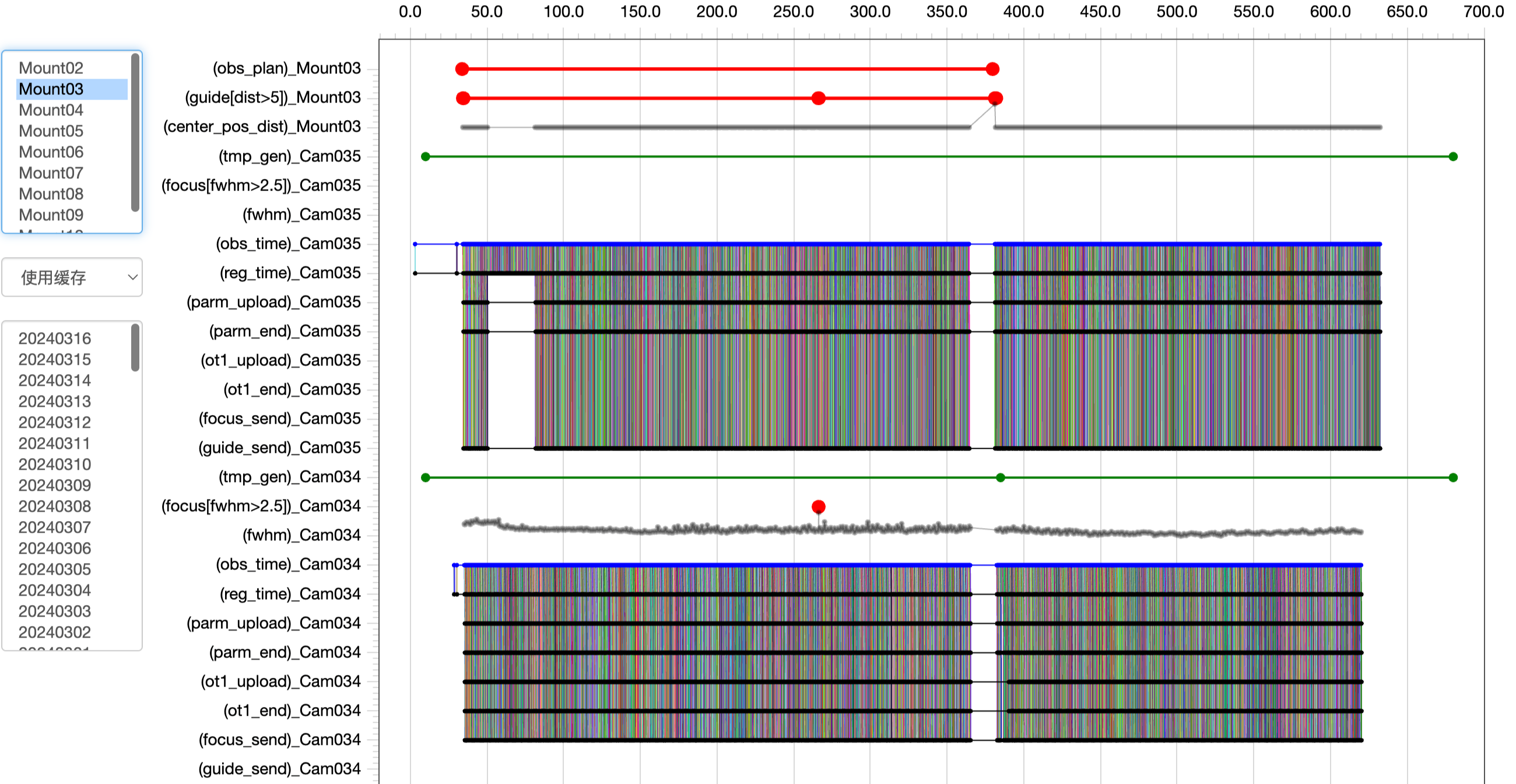}
\centering
\caption{State Evolution monitoring view. The X-axis represents the time difference relative to the start of the observation, with units in minutes. The Y-axis displays a list of key parameters related to the mount, camera, and image, arranged sequentially according to the cameras.
\label{fig:stateEvolution}}
\end{figure}

\subsubsection{Key Parameter Monitoring}

This view monitors parameters related to mounts, cameras, and images. Mount pointing accuracy, as shown in the left image of Figure 4, is plotted over time, showing deviations from planned pointing. Image quality (Figure 4, right), influenced by camera and environmental factors, is monitored via parameters like FWHM, star count, background brightness, limiting magnitude, and processing time. Camera hardware parameters (Figure 4, right), such as temperature, are also displayed over time.

\begin{figure}
\gridline{
  \includegraphics[height=4cm]{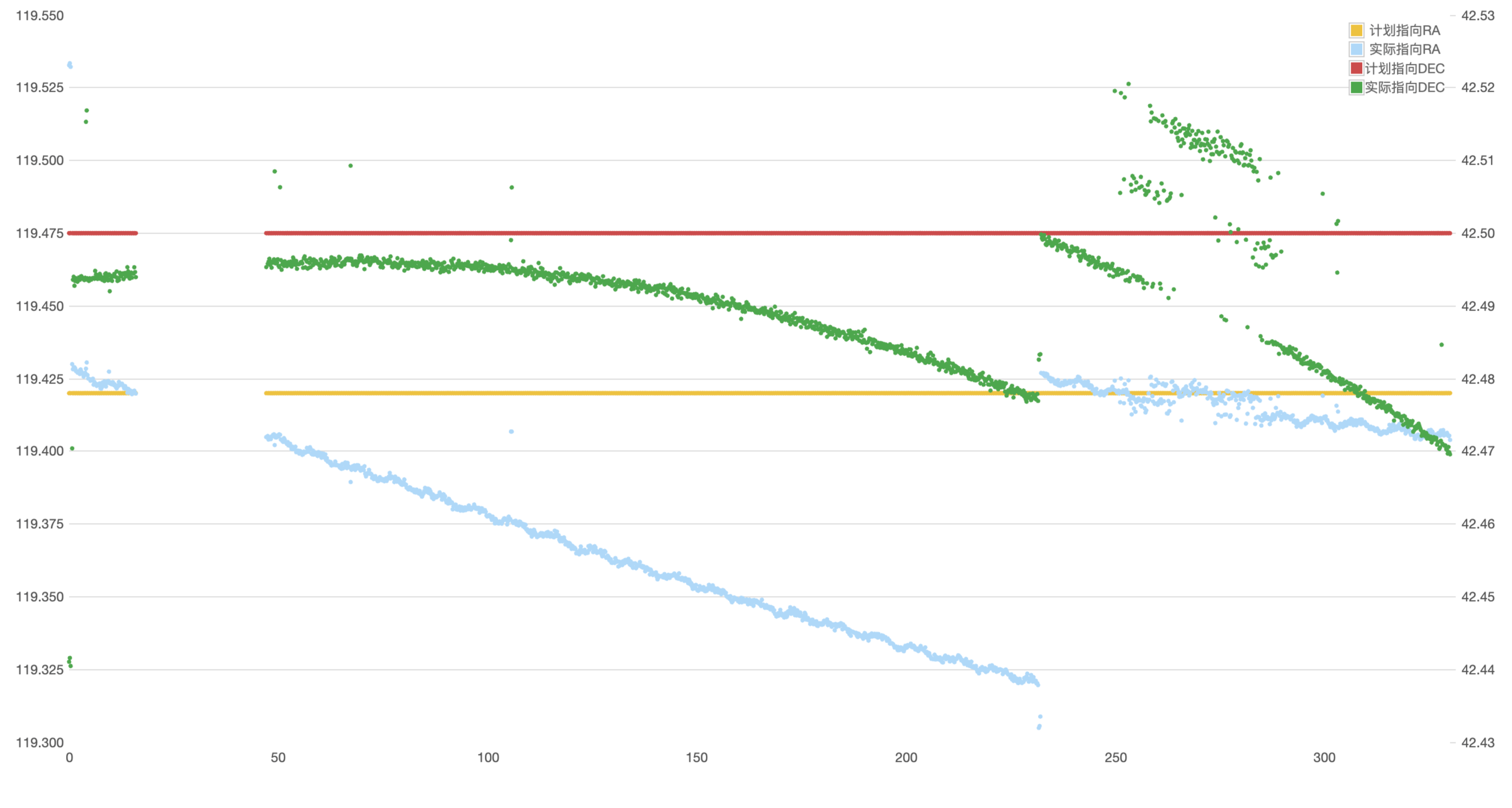}
  \includegraphics[height=4cm]{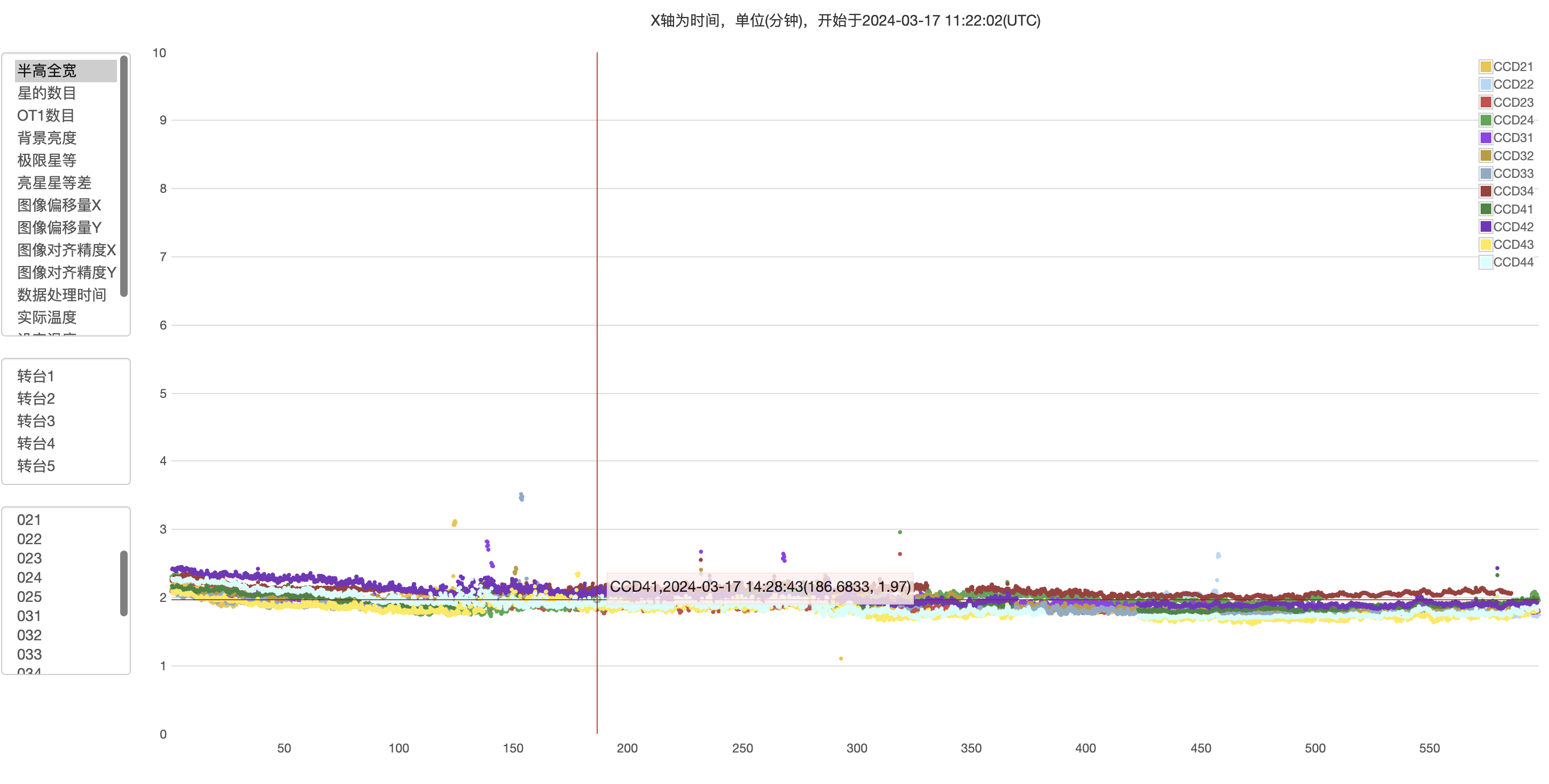}
}
\caption{Key parameter monitoring. The left image shows the mount parameters, while the right image displays the image parameters.\label{fig:keyParameter}}
\end{figure}

\subsubsection{Transient Lifecycle Monitoring}

Thanks to GWAC's hardware characteristics, it has the capability to observe transients with durations on the order of minutes. GWAC has completed the entire process from transient detection to automatic identification and has successfully observed a large number of minute-scale transients, such as the 200 white-light flares observed by \citep{2023RAA....23a5016L}. In this process, follow-up identification using the 60 cm telescope at the same site \citep{2020PASP..132e4502X} is a critical step. Early identification of transients helps to trigger large-aperture telescopes for spectroscopic and other follow-up observations sooner. Therefore, optimizing the data processing pipeline of GWAC to accelerate the process from detection to automatic identification of transients is particularly important.

To this end, we proposed the concept of the transient lifecycle, aimed at monitoring the key modules in the transient processing pipeline and optimizing accordingly. Key modules in the transient lifecycle include the detection, identification, and follow-up observation. The system records the time consumption at each module to monitor the transient lifecycle. As shown in Figure 5, the transient lifecycle monitoring view is depicted. The left side is the functional control area, including a list of key modules in the transient processing pipeline, a mount list, and a camera list. The right side shows a scatter plot of the time consumption at the selected key modules over time, illustrating the distribution of time consumption at a particular moment. If the system encounters a fault, there will be drastic changes in time consumption across stages, thus reflecting system anomalies in a timely manner. Figure 8 in Section 4 illustrates a fault about  transient lifecycle.

\begin{figure}
\centering
\includegraphics[height=5cm]{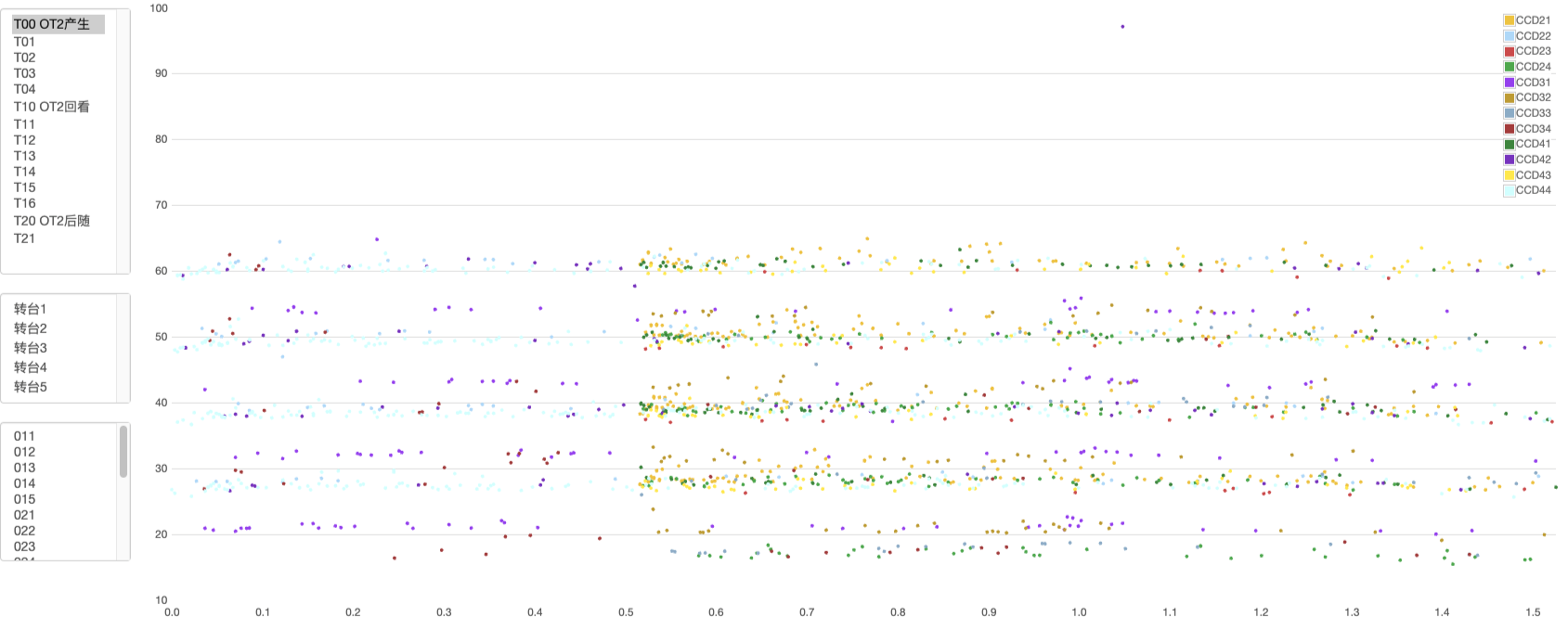}
\caption{Transient lifecycle monitoring. The time consumption at each modules in the transient lifecycle varies over time. The X-axis represents the start time of the key modules, in minutes, indicating the time difference between the start time of the key modules and the beginning of the observation. The Y-axis represents the time comsumeing of the key modules, in seconds, which started at this moment. \label{fig:transientLifecycle}}
\end{figure}

\subsection{Database Design}

To support data storage for the monitoring system, we have designed a series of database tables. These tables are divided into two main categories: Basic Entity Tables and Entity State Tables.
\begin{enumerate}
\item Basic Entity Tables: This includes the observation plan table, mount table, camera table, image table, and transient table. The basic entity tables are shared with the scientific data processing pipeline and are used to support the operation and result management of GWAC.
\item Entity State Tables: This consists of instantaneous state table, time consume table of transient key modules, image lifecycle monitoring table, image and instrument parameter table, etc. Those tables are specific to the monitoring system and are used to record the data parameters, state parameters, and the temporal changes of the lifecycle of the monitored entities.
\end{enumerate}

\subsection{System Implementation}

The GWAC monitoring system comprises a web server, a database, and several data collection programs. The web server facilitates the display of various views and provides API interfaces for data collection. 
These data collection programs run on the data processing servers of GWAC, where they monitor and gather local operational state information.
This information is uploaded to the web server via the API interface during each image cycle. The collected monitoring data is then stored in the database, allowing users to access it through multiple view pages on the web browser.

\section{Fault Diagnosis Practices Based on Monitoring Views} \label{sec:faultDiagnosis}

During the one-year operation of the monitoring system, extensive fault analysis and diagnostics were conducted based on the monitoring views, leading to the establishment of a correspondence table between faults and monitoring views. Almost all common faults have a corresponding view and solution within this table. With these monitoring views, the process from fault warning to analysis and diagnosis typically takes only a few minutes. Compared to traditional methods, which could take anywhere from several tens of minutes to hours, this solution has improved the speed of fault diagnosis by more than tenfold. It should be noted that the current monitoring views do not yet cover all possible fault information, and the system is undergoing continuous upgrades and optimization. In the future, the monitoring content will be further expanded to cover more fault points as comprehensively as possible.

The following are examples of common fault diagnoses based on monitoring views. 
\begin{enumerate}
\item Defocused Image: 
The left image in Figure 6 shows a real-time preview of the observed image. The poor image quality indicates a failure in the autofocus function.
\item Response Timeout of Single or Multiple Module : 
The right image in Figure 6 shows the instantaneous status monitoring of key modules. Each rectangle in the image represents a module, with the majority of rectangles typically colored green. A red rectangle indicates an abnormality in the corresponding module. If an entire column is red, it means that the module of all cameras has either failed to start or has encountered a complete failure.
\item Data Processing Failed Simultaneously in Multiple Nodes: 
Figure 7 presents three cases of State Evolution Monitoring. 
In the left image of Figure 7, the FWHM of two cameras is unusually large due to weather conditions, which caused a degradation in image quality and led to simultaneous failures of data processing on two nodes of the same mount. 
The middle image of Figure 7 displays multiple failures. First, a weather-related fault caused data processing failures on all nodes of one mount, triggering a sky region switch. Frequent sky region changes deplete the observation plan for that time period, eventually causing the mount to enter a waiting state and stop observations. Failures in the focus of guiding camera or excessive pointing deviation can also lead to simultaneous data processing failures across all nodes of a mount. Additionally, a failure in one node's SExtractor caused images to be captured but not processed.
\item Data Processing Delay of Single Node: 
The right image in Figure 7 shows a network card failure in the camera control server, leading to a transmission speed lower than the image readout rate, which caused significant delays in image transmission. However, the processing in subsequent modules continued normally.
\item Anomalous Transient Lifecycle: 
Figure 8 shows the time distribution of transient candidate catalog parsing. There are two time periods in which the parsing of the transient catalog experienced delays. This is typically caused by consecutive processing failures of multiple images, which generate an excessive number of transient candidates. When the number of candidates exceeds the processing peak capacity of the pipeline, they begin to queue, resulting in processing delays. As the processing continues, the number of queued targets gradually decreases, and the processing time returns to normal.

\end{enumerate}

\begin{figure}
\gridline{
  \includegraphics[height=4cm]{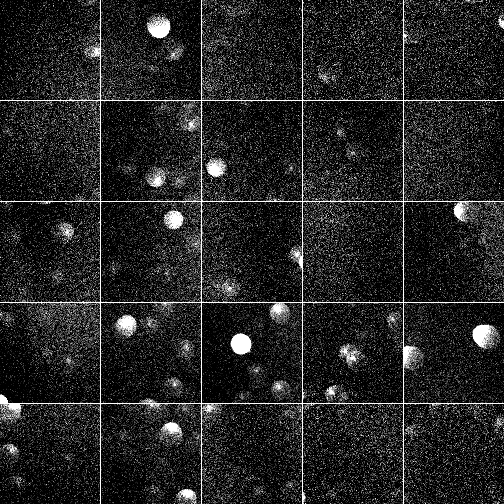}
  \includegraphics[height=4cm]{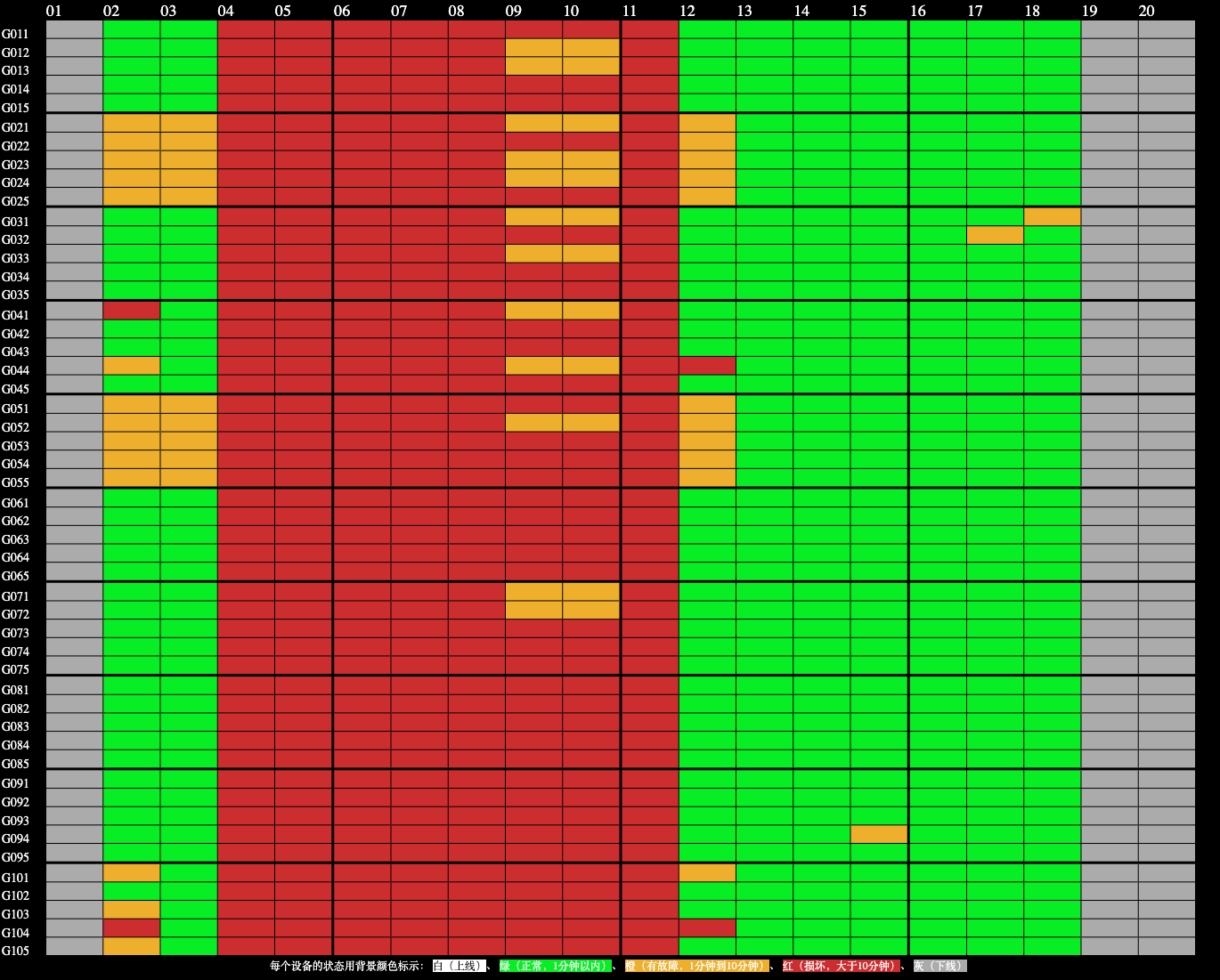}
}
\caption{Instantaneous status fault example. The left image shows a defocused image, while each red square in the right image represents the anomaly of a module.\label{fig:instantaneousExample}}
\end{figure}

\begin{figure}
\gridline{
  \includegraphics[height=4cm]{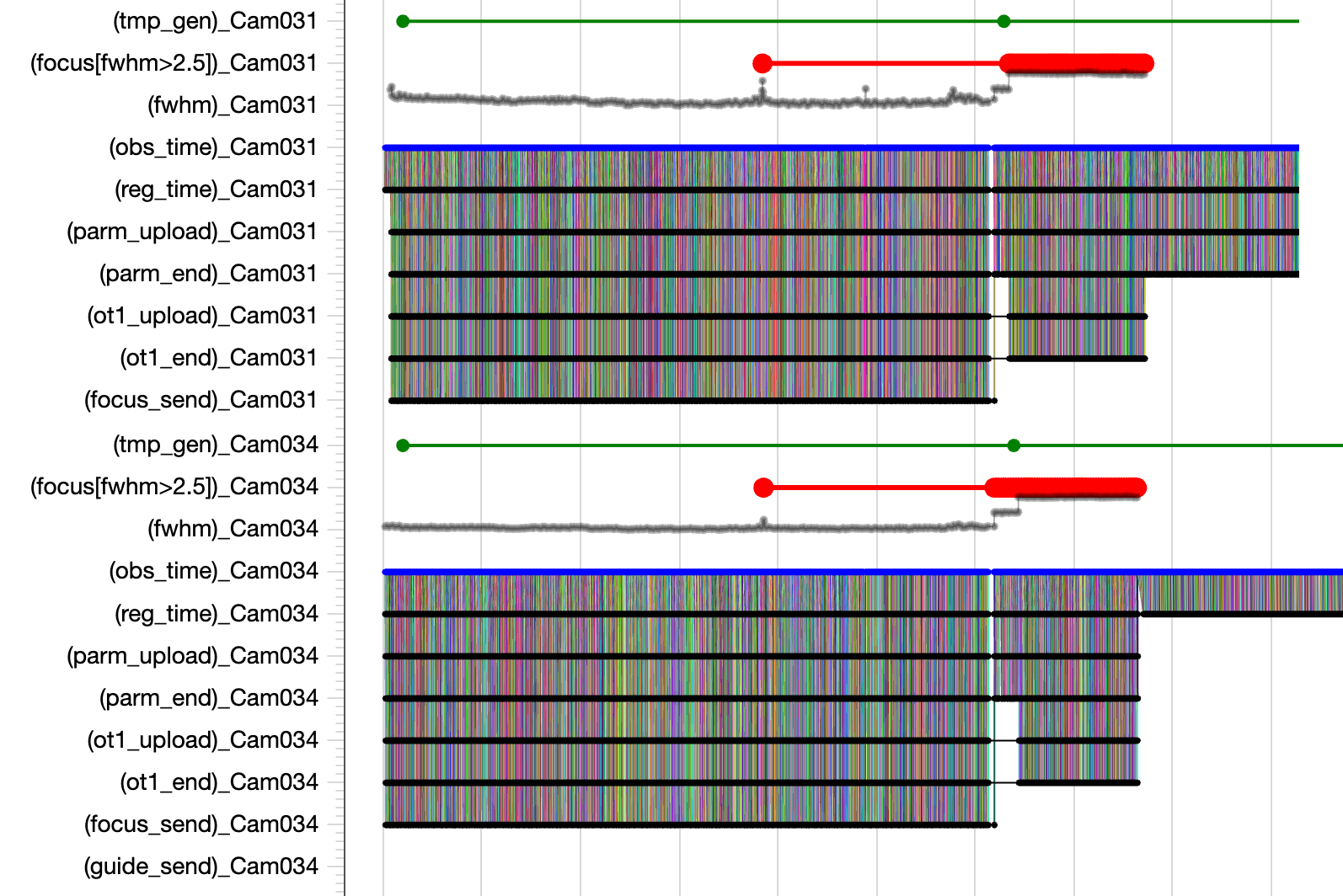}
  \includegraphics[height=4cm]{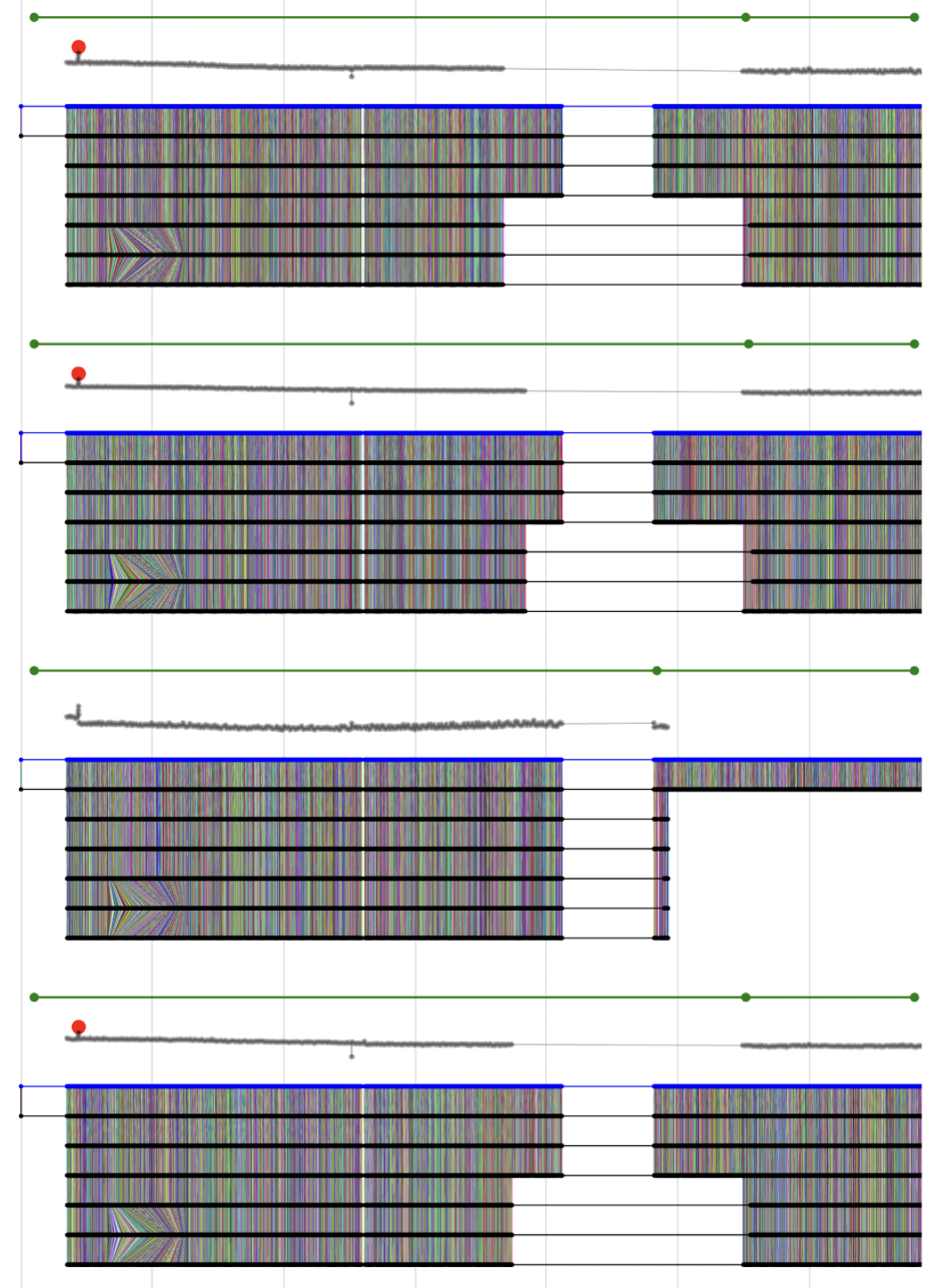}
  \includegraphics[height=4cm]{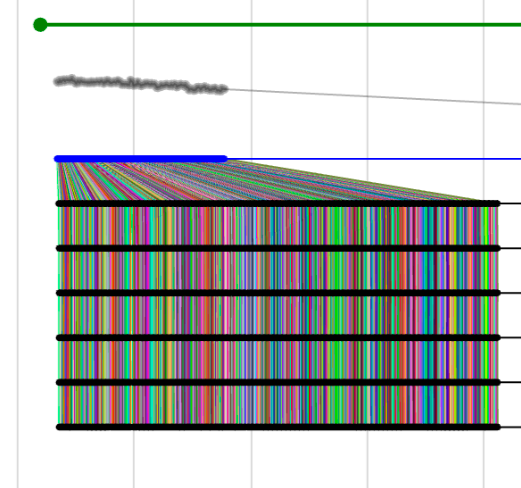}
}
\caption{Three example of State Evolution monitoring\label{fig:stateEvolutionExample}}
\end{figure}

\begin{figure}
\centering
\includegraphics[height=7cm]{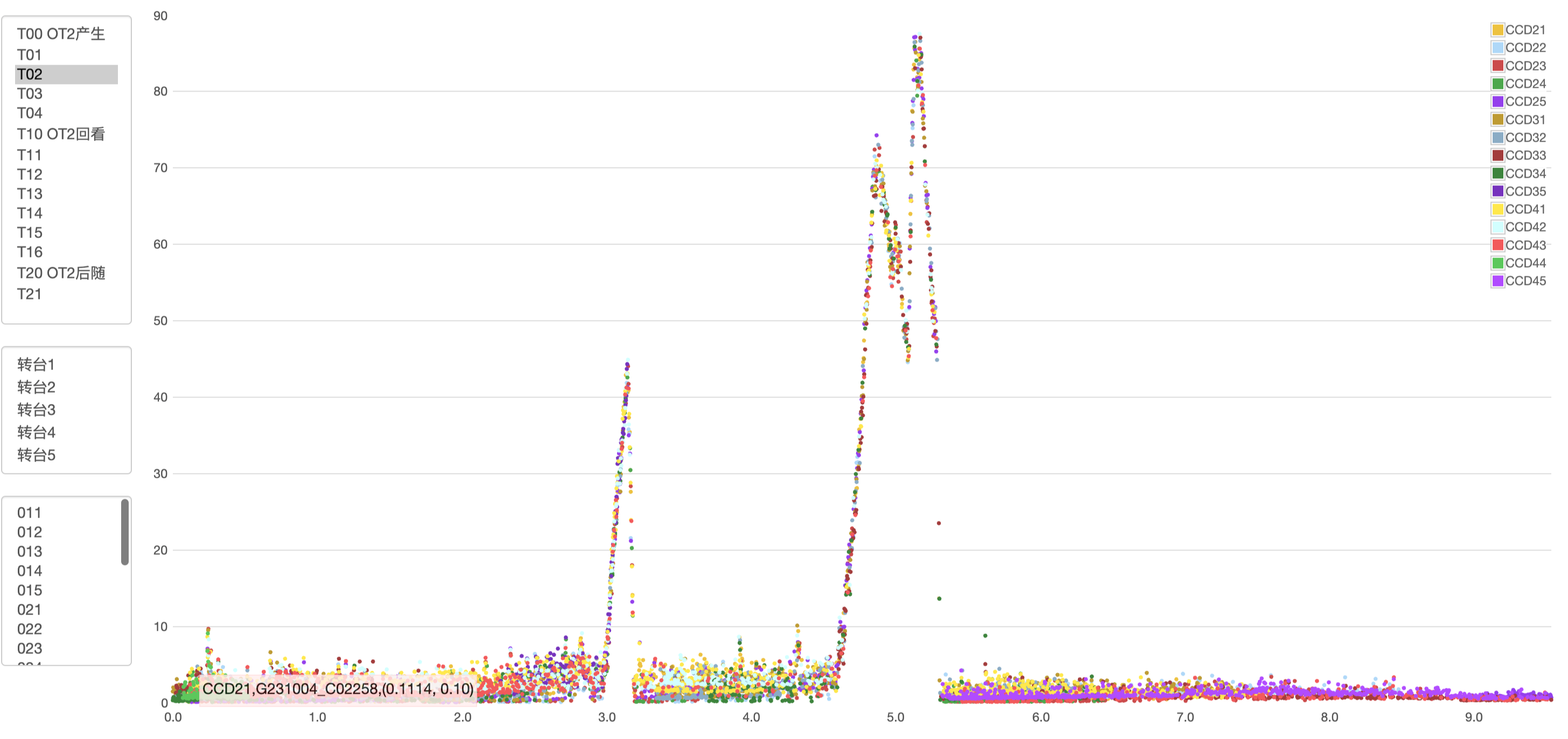}
\caption{Example of transient lifecycle fault. An excessive number of transient candidates causes congestion in the identification stage.\label{fig:ot2time}}
\end{figure}

\section{Summary and Future Work} \label{sec:summary}

The data processing pipeline of GWAC is complex, resulting in a high failure rate and making fault diagnosis difficult. To accelerate the efficiency of fault diagnosis, this paper designs and implements a monitoring system.
The primary innovation lies in the design of monitoring views tailored to the complex pipeline, which abstracts the intricate data processing of over 2,500 software modules across 50 telescopes and presents this information in a concise and efficient manner on limited computer screens. The system features four views: instantaneous status monitoring, State Evolution monitoring, key parameter monitoring, and transient lifecycle monitoring, providing a comprehensive view of the monitoring data. Notably, the state evolution monitoring and transient lifecycle monitoring are monitoring views proposed for the first time in the astronomical field. 

Through these monitoring views, users gain comprehensive insights into the instantaneous states and state evolution of GWAC, including pointing, observation, data processing, and software/hardware feedback. This capability significantly accelerates fault detection and localization. Additionally, the lifecycle monitoring of transients provides a clear visualization of critical anomalies at key process modules, serving as valuable reference indicators for optimizing observation efficiency. Practical case analysis shows that the fault localization speed of GWAC improved by nearly tenfold with the system. The paper also details the monitoring data collection and storage schemes. Finally, several monitoring case studies are presented  to illustrate the fault analysis based on different monitoring views. 

The current system primarily focuses on the visualization of monitoring data, aiding staff in comprehending the system's operational status and conducting fault analysis. While simple faults can be diagnosed directly through the monitoring views, complex issues require expert intervention using the views alongside detailed backend logs. Future work will involve compiling fault cases and developing a knowledge base linking monitoring data to faults. Additionally, research will be conducted on an automated fault diagnosis system based on monitoring data, aiming to further lower the fault diagnosis threshold and improve system maintenance efficiency. The design of the system's monitoring views exhibits high generalizability, making it applicable not only to GWAC but also to the status monitoring of other camera arrays.

\section{Acknowledgement}

I would like to thank the members of the GWAC team for their suggestions and help during the experiment. 
This paper was supported by the Young Data Scientist Program of the China National Astronomical Data Center, the Strategic Priority Research Program of the Chinese Academy of Sciences (Grant No. XDB0550401), and the National Natural Science Foundation of China (Grant No. 12494573).


\begin{thebibliography}{}

\bibitem[Wei et al.(2016)]{2016arXiv161006892W} Wei, J., Cordier, B., Antier, S., et al.\ 2016, arXiv e-prints, arXiv:1610.06892
\bibitem[Xin et al.(2023)]{2023NatAs...7..724X} Xin, L., Han, X., Li, H., et al.\ 2023, Nature Astronomy, 7, 724. doi:10.1038/s41550-023-01930-0
\bibitem[Li et al.(2024)]{2024ApJ...971..114L} Li, G.-W., Wang, L., Yuan, H.-L., et al.\ 2024, \apj, 971, 114. doi:10.3847/1538-4357/ad55e8
\bibitem[Li et al.(2023)]{2023RAA....23a5016L} Li, G.-W., Wu, C., Zhou, G.-P., et al.\ 2023, Research in Astronomy and Astrophysics, 23, 015016. doi:10.1088/1674-4527/aca506
\bibitem[Xin et al.(2024)]{2024MNRAS.527.2232X} Xin, L.-P., Li, H.-. li ., Wang, J., et al.\ 2024, \mnras, 527, 2232. doi:10.1093/mnras/stad960
\bibitem[Xin et al.(2021)]{2021ApJ...909..106X} Xin, L.~P., Li, H.~L., Wang, J., et al.\ 2021, \apj, 909, 106. doi:10.3847/1538-4357/abdd1b
\bibitem[Li et al.(2023)]{2023ApJ...954..142L} Li, H.-L., Wang, J., Xin, L.-P., et al.\ 2023, \apj, 954, 142. doi:10.3847/1538-4357/ace59b
\bibitem[Wang et al.(2022)]{2022ApJ...934...98W} Wang, J., Li, H.~L., Xin, L.~P., et al.\ 2022, \apj, 934, 98. doi:10.3847/1538-4357/ac7a35
\bibitem[Han et al.(2021)]{2021PASP..133f5001H} Han, X., Xiao, Y., Zhang, P., et al.\ 2021, \pasp, 133, 065001. doi:10.1088/1538-3873/abfb4e
\bibitem[Xu et al.(2020)]{2020PASP..132e4502X} Xu, Y., Xin, L.~P., Wang, J., et al.\ 2020, \pasp, 132, 054502. doi:10.1088/1538-3873/ab7a73
\bibitem[Xu et al.(2021)]{2021RMxAC..53..174X} Xu, Y., Xin, L.~P., Han, X.~H., et al.\ 2021, Revista Mexicana de Astronomia y Astrofisica Conference Series, 53, 174. doi:10.22201/ia.14052059p.2021.53.36
\bibitem[Costa et al.(2022)]{2022icrc.confE.700C} Costa, A., Munari, K., Incardona, F., et al.\ 2022, 37th International Cosmic Ray Conference, 700. doi:10.22323/1.395.0195
\bibitem[Hu et al.(2021)]{2021MNRAS.500..388H} Hu, T.~Z., Zhang, Y., Cui, X.~Q., et al.\ 2021, \mnras, 500, 388. doi:10.1093/mnras/staa3087
\bibitem[Di Carlo et al.(2016)]{2016SPIE.9913E..3SD} Di Carlo, M., Dolci, M., Smareglia, R., et al.\ 2016, \procspie, 9913, 99133S. doi:10.1117/12.2231614



\end{thebibliography}
\end{document}